%% file: yrc.tex
\documentclass[aps,prd,twocolumn,showpacs,superscriptaddress,groupedaddress]{revtex4}
\usepackage{graphicx}
\usepackage{epsfig}
\usepackage{dcolumn}
\usepackage{bm}
\usepackage[usenames]{color} 
\usepackage{subfigure}
\usepackage{amsmath}
\usepackage{amssymb}
\usepackage{float}
\usepackage[colorlinks=true, pdfstartview=FitV, linkcolor=blue, citecolor=blue, urlcolor=blue]{hyperref}

\begin{document}

\hspace{5.2in} \mbox{Fermilab-Pub-13-405-E}

\title{Search for the {\boldmath $X(4140)$} state in {\boldmath  $B^+ \rightarrow J/\psi \phi K^+$} decays with the D0 detector}

\input author_list.tex  
\date{September 25,2013}

\begin{abstract}

We investigate the decay $B^+ \rightarrow J/\psi \phi K^+$  in a search for the $X(4140)$ state, 
 a narrow threshold resonance in the  $J/\psi \phi$  system. The data sample corresponds to an integrated
 luminosity of  $10.4~\rm{fb^{-1}}$ of $ p \overline p $ collisions at $\sqrt s =1.96$~TeV collected by the 
D0 experiment at the Fermilab Tevatron collider. We observe a mass peak with a statistical significance of 3.1 standard deviations and measure its
 invariant mass to be $M=4159.0\pm4.3 {\rm \thinspace (stat)} \pm 6.6 {\rm \thinspace (syst)}$~MeV and
 its width to be $\Gamma= 19.9\pm12.6 {\rm \thinspace (stat)}^{+1.0}_{-8.0} {\rm \thinspace (syst)}$~MeV. 
Identifying this mass peak as the $X(4140)$, we measure,  for $M(J/\psi \phi) <4.59$~GeV,
the relative branching fraction ${\cal B}_{\rm rel}= {\cal B}(B^+ \rightarrow X(4140) K^+)/ {\cal B}(B^+ \rightarrow J/\psi \phi K^+)$
to be $21 \pm 8 {\rm \thinspace (stat)} \pm 4 {\rm \thinspace (syst)}$\%.  In addition, the data can accommodate the presence of a second enhancement at a mass of $4328.5\pm 12.0$~MeV.

\end{abstract}


\maketitle





The $X(4140)$ state~\cite{pdg2012} is a narrow resonance in the $J/\psi \phi$ system produced near threshold.  The CDF Collaboration reported the first 
evidence~\cite{Aaltonen:2009at} for this state (termed $Y(4140)$) in the decay $B^+ \rightarrow J/\psi \phi K^+$ (charge conjugation is implied throughout) and measured the invariant mass $M=4143.0 \pm2.9 (\rm {stat}) \pm 1.2 {\rm \thinspace (syst)}$~MeV and width $\Gamma= 11.7^{+8.3}_{-5.0} {\rm \thinspace (stat)} \pm 3.7 {\rm \thinspace (syst)}$~MeV.  

 The Belle Collaboration searched for $X(4140)$ in the process $\gamma \gamma \rightarrow J/\psi \phi$ and, finding no significant signal, reported upper limits on the product of
the partial width $\Gamma_{\gamma \gamma}$ and branching fraction $X(4140)\rightarrow J/\psi \phi$
for $J^P =0^+$ and $2^+$~\cite{Shen:2009vs}.  
 At the LHC, both  the LHCb and CMS Collaborations have searched for the state. The LHCb Collaboration found no evidence~\cite{Aaij:2012pz},
 in disagreement with the CDF measurement.  A preliminary report~\cite{cms} from the CMS Collaboration on a search for the same signature supports the 
CDF observation.  With two out of four experiments failing to observe the $X(4140)$ resonance the question of the existence of this state still remains open.  A detailed review is given in Ref.~\cite{ky}.

The quark model of three-quark baryons and quark-antiquark mesons does not predict a hadronic state at this mass. The decay channel suggests that this resonance may be a $c \overline c $ bound state. 
However, at this mass, above the open-charm threshold of 3740 MeV, it is unlikely to be a conventional charmonium state. Such states  are expected to
 decay predominantly to pairs of charmed mesons and they would have a much larger width than experimentally observed. It has been suggested that $X(4140)$ is a molecular 
structure made of two charmed mesons, e.g. $(D_s,\overline D_s)$, but  other possible states  are hybrid particles composed of two quarks and a 
valence gluon ($q\overline qg$) or four-quark combinations ($c\overline c s \overline s$). 
For details see the review of hadronic spectroscopy in Ref.~\cite{Drenska:2010kg} and references therein.

In addition to $X(4140)$, the CDF Collaboration reported seing a second enhancement in the same channel, located near 4.29 GeV.
 A similar structure is also seen by the CMS Collaboration~\cite{cms}.
 Belle also reports a new structure
at $M=4350.6^{+4.6}_{-5.1}{\rm \thinspace (stat)}\pm0.7  {\rm \thinspace (syst)}$~MeV.

In this Article we present results of a search for the $X(4140)$ resonance and any excited states in the  $J/\psi \phi$ system  in the decay sequence $B^+ \rightarrow J/\psi \phi K^+$, $J/\psi \rightarrow \mu^+ \mu^-$, $\phi \rightarrow K^+K^-$. The data sample corresponds to an integrated luminosity of 10.4 fb$^{-1}$ collected with the D0 detector  in $p \overline p $ collisions at the Fermilab Tevatron collider.

The D0 detector consists of a central tracking system, calorimetry system and
muon detectors, as detailed in Ref.~\cite{Abazov2006463}. The central
tracking system comprises  a silicon microstrip tracker (SMT) and a central
fiber tracker (CFT), both located inside a 1.9~T superconducting solenoidal
magnet.  The tracking system is designed to optimize tracking and vertexing
for pseudorapidities $|\eta|<3$,
where  $\eta = -\ln[\tan(\theta/2)]$, and  $\theta$ is the 
polar angle with respect to the proton beam direction.
  The SMT can reconstruct the $p\overline{p}$ interaction vertex (PV) 
for interactions   with at least three tracks with a precision
of 40~$\mu$m in the plane transverse to the beam direction and determine
the impact parameter of a track relative to the PV with a precision between
20 and 50 $\mu$m, depending on the number of hits in the SMT.
The muon detector, positioned outside the calorimeter, consists of a central muon system covering the pseudorapidity region of $|\eta|<1$ and a forward muon system covering the pseudorapidity region of $1<|\eta|<2$. Both central and forward systems consist of a layer of drift  tubes
and scintillators inside 1.8~T toroidal magnets and two similar layers outside the toroids~\cite{d0mu}.

We use the Monte Carlo (MC) event generator {\sc pythia}~\cite{pythia}  interfaced with the particle decay package {\sc EvtGen}~\cite{evtgen} to simulate the
decay chain $B^+ \rightarrow J/\psi \phi K^+$, $J/\psi \rightarrow \mu^+ \mu^-$, $\phi \rightarrow K^+K^-$.
The $B^+$ decay is simulated according to  three-body phase space.
The detector response is simulated with {\sc geant}~\cite{geant}. Simulated signal events are
overlayed with events from randomly
collected $p \overline p$ bunch crossings to simulate multiple interactions.

Events used in this analysis are collected with both single-muon and dimuon triggers.
Candidate events are required to include a pair of  oppositely charged muons accompanied by three additional charged particles with transverse momenta above 0.7~GeV.  Both muons are required to be detected in the muon chambers inside the toroid magnet, and at least one of the muons is required to be also detected outside the toroid~\cite{muid}.
 Each of the five final-state tracks is required to have at least one SMT hit and at least one CFT hit.

To form $B^+$ candidates, muon pairs in the invariant mass range $2.9<M(\mu^+ \mu^-)<3.3$~GeV, consistent with $J/\psi$ decay, are combined with pairs of oppositely charged particles (assigned the kaon mass hypothesis) with an invariant mass in the range $0.99 <M(K^+K^-)<1.07$~GeV and with a third charged particle, also assigned the kaon mass hypothesis. The third kaon is required to have at least three SMT hits.
The dimuon invariant mass is constrained in the kinematic fit to the world-average   $J/\psi$ mass~\cite{pdg2012}
and   the five-track system is constrained to a common vertex.
 The trajectories of the five $B^+$ decay products are adjusted according to the decay and  kinematic fit.
The adjusted track parameters are used in the
calculation of  the $B^+$ candidate mass. 
The $B^+$ candidates are required to
have  an invariant  mass in the range $5.15 <M(J/\psi K^+K^-K^+)<5.45$~GeV.
The  $\chi^2$ of the $B^+$ vertex fit is required to be less than 20 for 6 degrees of freedom, with the contribution of the third kaon to the $\chi^2$ required to be less than 4.

To reconstruct the PV, tracks are selected that 
do not originate from the candidate $B^+$ decay, 
and a constraint is applied to the average beam-spot position in the transverse plane.
We define the signed  decay length of a $B^+$ meson, $L^B_{xy}$, 
as the  vector pointing
from the PV to the decay vertex, projected on the transverse plane.  We require $L^B_{xy}$ to be greater than 250 $\mu$m to suppress background from prompt $J/\psi$ production. The angle between the pointing vector and the $B^+$ meson transverse momentum is required to be less than 3.6$^\circ$. We also reconstruct the decay vertex of the $J/\psi \phi$ pair and require the distance between the $B^+$ and $J/\psi \phi$  vertices in the transverse plane and in the beam direction to be less than 50 $\mu$m and less than 150 $\mu$m, respectively (five  times the $RMS$ determined by MC).  The selection is limited to events with $M(J/\psi \phi)$ below 4.59~GeV. At larger masses background from other $b$ hadron decays is large.

Background arises from a misidentified $\phi$ meson or a misidentified third kaon. To suppress background contribution from combinations including particles produced in the hadronization process or in other $B$ hadron decays, we require the transverse momentum of the  $B^+$ meson to be between 7 and 30 GeV. The fraction of the $B^+$ transverse momentum carried by the three kaons is required to be greater than 0.2. We remove decays $B \rightarrow \psi(2S) +X$ by vetoing the  mass range
$3.661<M(J/\psi \pi^+ \pi^-)<3. 711$~GeV, equivalent to $\pm 2.5$ standard deviations around the world-average  $\psi(2S)$ mass~\cite{pdg2012}, for all combinations of $J/\psi$ produced with a pair of oppositely charged
particles assigned a pion mass hypothesis. For the remaining sample, we accept one candidate per event, selecting
the combination with the lower  $\phi$ candidate mass. Simulations show that this choice is 95\% efficient for the signal. 
Any sample bias  resulting from the above selection is quantified and corrected using the  efficiency determined by MC simulations.

\begin{figure}
  \centering
   \includegraphics[width=0.80\columnwidth]{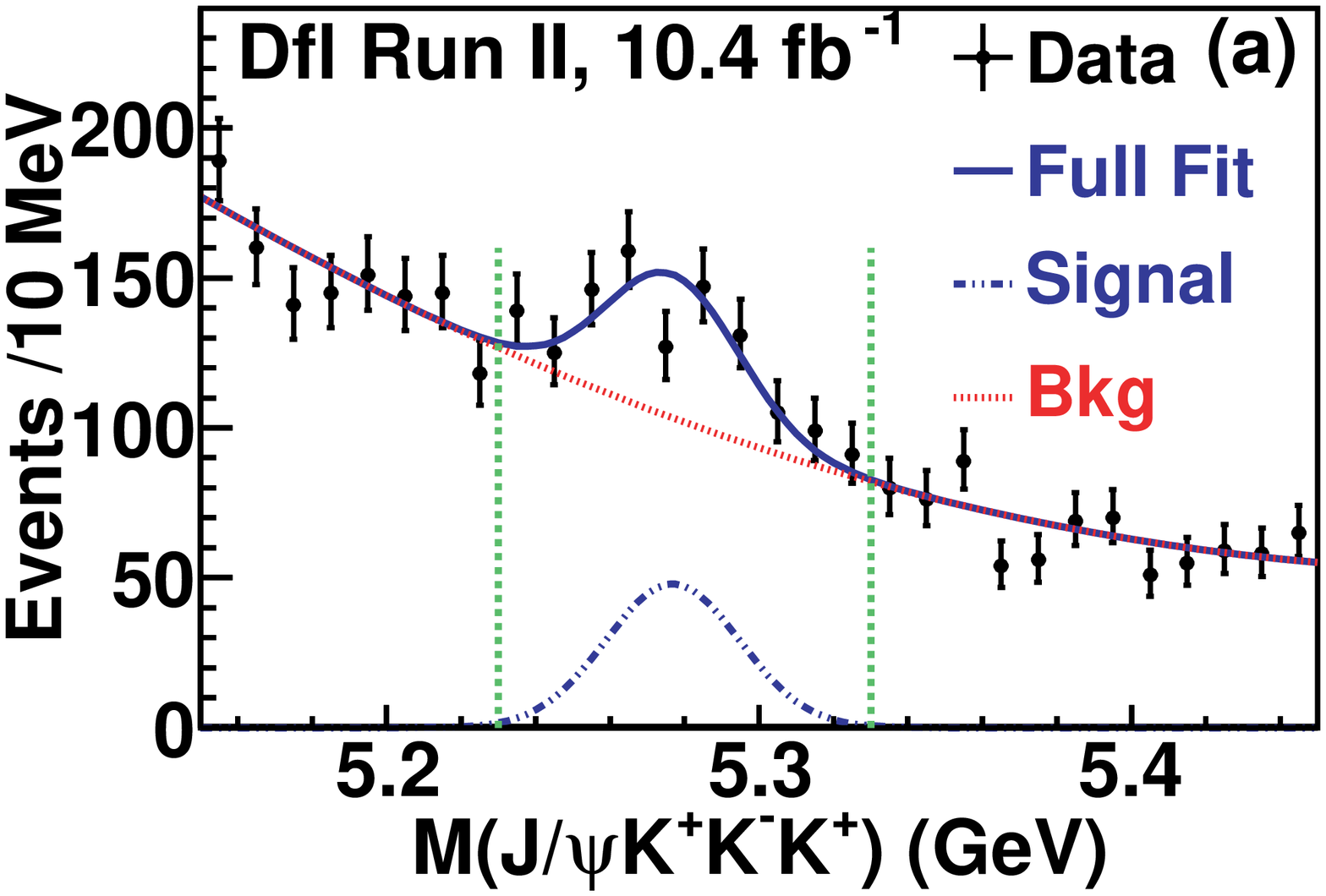}
   \includegraphics[width=0.80\columnwidth]{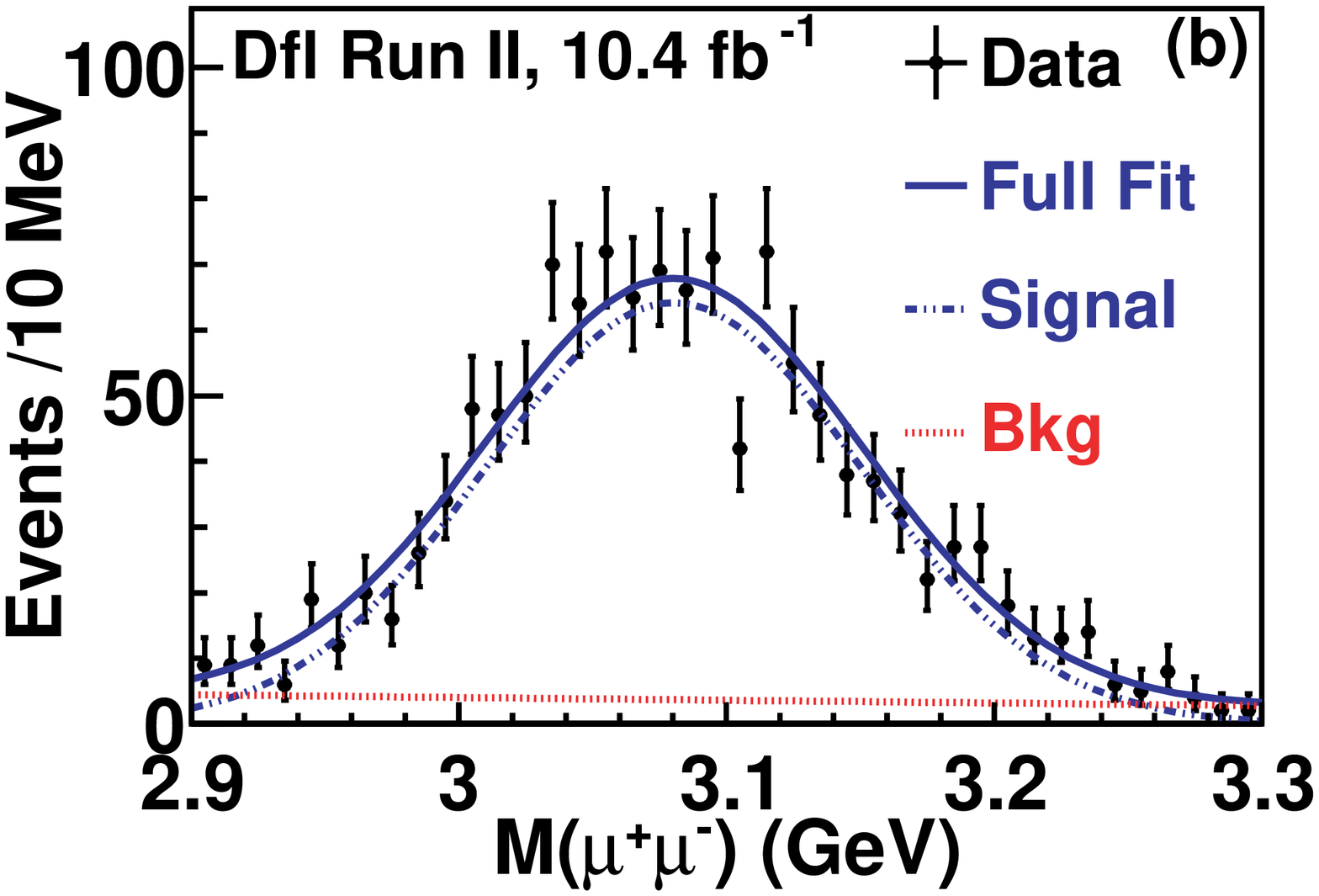}
  \caption{(color online) (a) Invariant mass distribution of $B^+ \rightarrow J/\psi \phi K^+$ candidates after the $1.005 < M(\phi) < 1.035$~GeV requirement. The fit  of a Gaussian signal with a second-order Chebyshev polynomial    background  is superimposed. (b) Invariant mass distribution $M(\mu^+ \mu^-)$ after the $B^+$ and $\phi$ mass window requirements.
 The fit  of a Gaussian function with a second-order Chebyshev polynomial  background  is superimposed.
The vertical green lines define the $B^+$ signal region.}
   \label{fig:ball}
\end{figure}

The $J/\psi \phi K^+$ invariant mass distribution for $B^+$ decay candidates  satisfying the mass requirement   $1.005 < M(\phi) < 1.035$~GeV consistent with the $\phi$ mass is shown in Fig.~\ref{fig:ball}(a). A binned maximum-likelihood fit  of a Gaussian signal with a mass resolution set to the value of 18 MeV (obtained from simulations), with a second-order Chebyshev polynomial  background, yields $215\pm37$ $B^+$ events with a mean mass of $M(B^+)=5277.8 \pm 3.3$~MeV, consistent with the world-average 
value of $B^+$ mass~\cite{pdg2012}. We define the signal mass range as $5.23 <M(B^+)<5.33$~GeV.  Figure~\ref{fig:ball}(b) shows the $J/\psi$ signal for events in the $B^+$ signal region. A fit of a Gaussian function and a second-order Chebyshev polynomial   background yields $1124\pm70$ $J/\psi$ events out of a total of 1269 $\mu^+ \mu^-$ candidates, showing that most of the selected events, including background, have a $J/\psi$ in the final state.

\begin{figure}[h!tb]
  \centering
   \includegraphics[width=0.80\columnwidth]{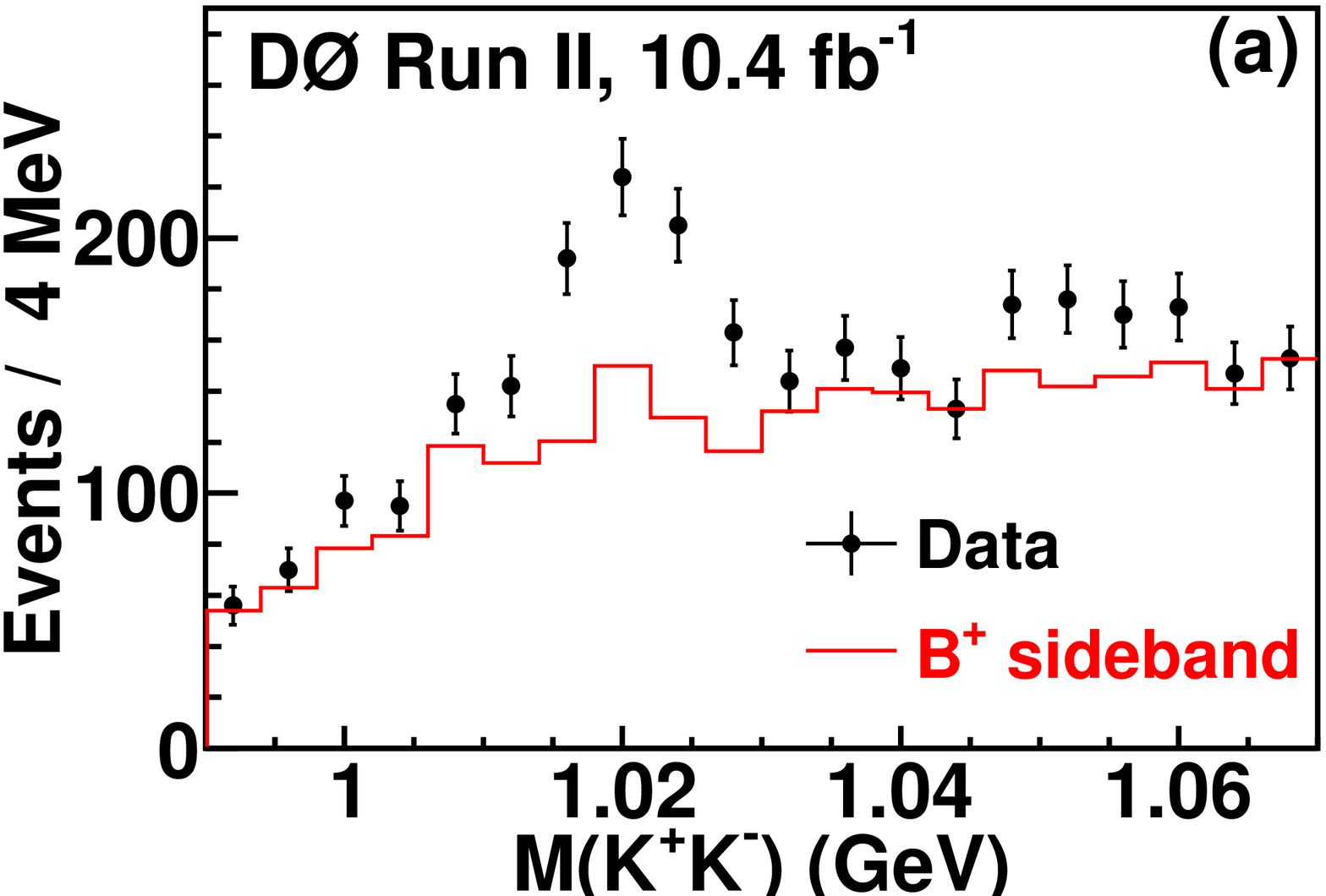}
   \includegraphics[width=0.80\columnwidth]{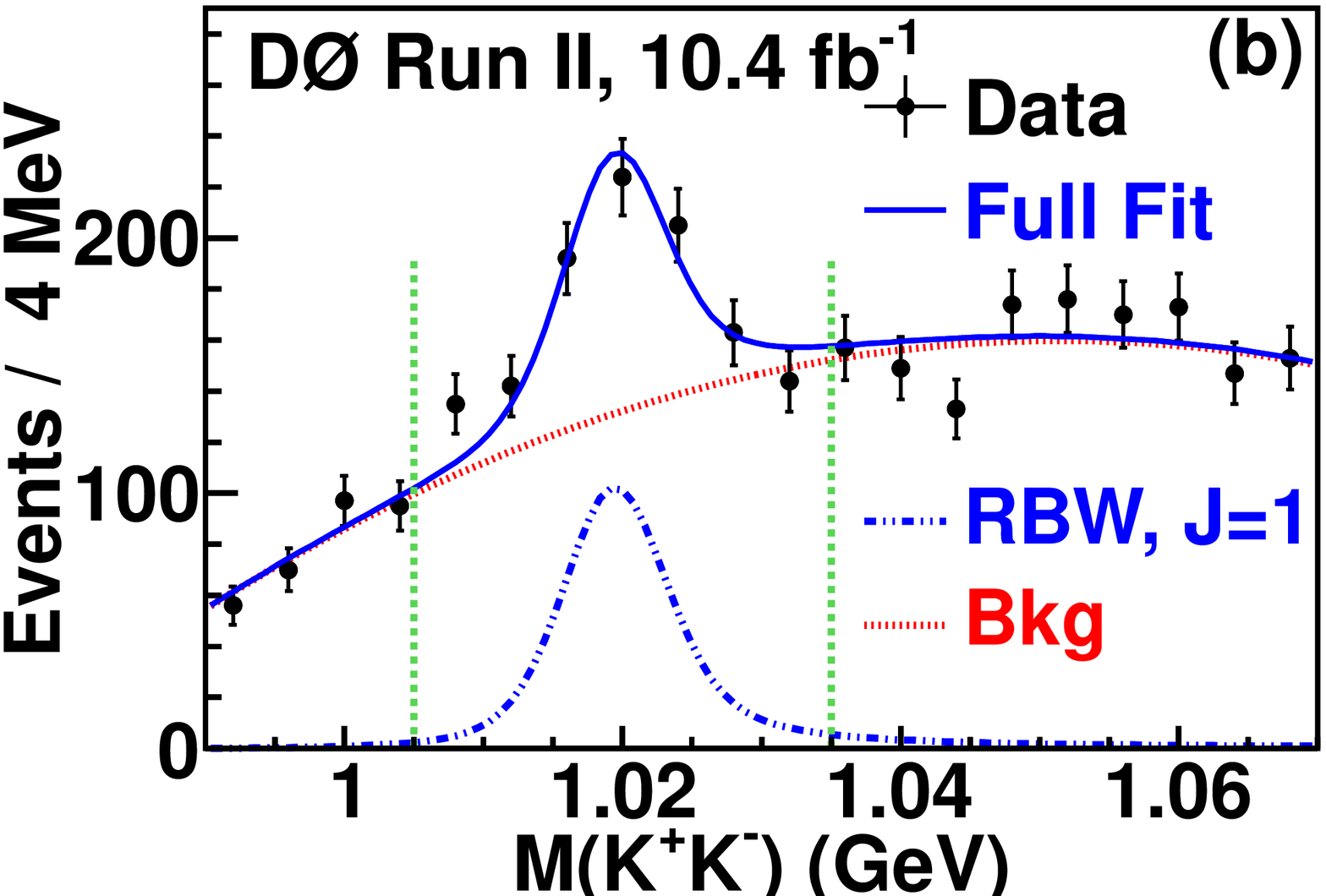}
  \caption{(color online) (a) Invariant mass distribution of $\phi$ candidates after the $B^+$ mass requirement $5.23<M(B^+)<5.33$ GeV  and in  the $B^+$ sidebands. (b)  Invariant mass distribution of $\phi$ candidates after the $B^+$ mass requirement. The fit of a relativistic $P$-wave Breit-Wigner function (RBW) with the world-average width of 4.26~MeV, convoluted with a Gaussian resolution of 3~MeV taken from simulations, with a second-order Chebyshev polynomial   background is superimposed.  The vertical green lines define the $\phi$ signal region
}
   \label{fig:kk}
\end{figure}

To establish a correspondence between the $B^+$ signal and the $\phi \rightarrow K^+K^- $ decay, we compare the invariant mass distributions of the $\phi$ candidates in the $B^+$ signal region and in the sidebands, defined as $5.15<M(J/\psi \phi K^+)<5.23$ GeV or $5.33<M(J/\psi \phi K^+)<5.45$~GeV. As seen in Fig.~\ref{fig:kk}(a),   there is a clear $\phi$ signal in the $B^+$ signal region, while the $\phi$ signal is much less pronounced in the $B^+$ sidebands. A fit, shown in Fig.~\ref{fig:kk}(b),  of a relativistic $P$-wave Breit-Wigner function with parameters set to the world-average values and a resolution of 3~MeV taken from simulations,  together with a second-order Chebyshev polynomial background, yields $284\pm40$ $\phi$ candidates. A similar fit to the $M(K^+K^-)$ distribution in the $B^+$ sideband yields $115\pm51$ $\phi$ candidates. Scaling the $\phi$ yield to the signal region leads to approximately 50 candidates. Thus, the total number of $\phi$ events in the $B^+$ signal region is consistent with the sum of the number of $B^+$ events determined in Fig.~\ref{fig:ball} and the expected contribution from background processes.

\begin{figure}[h!tb]
  \centering
   \includegraphics[width=0.80\columnwidth]{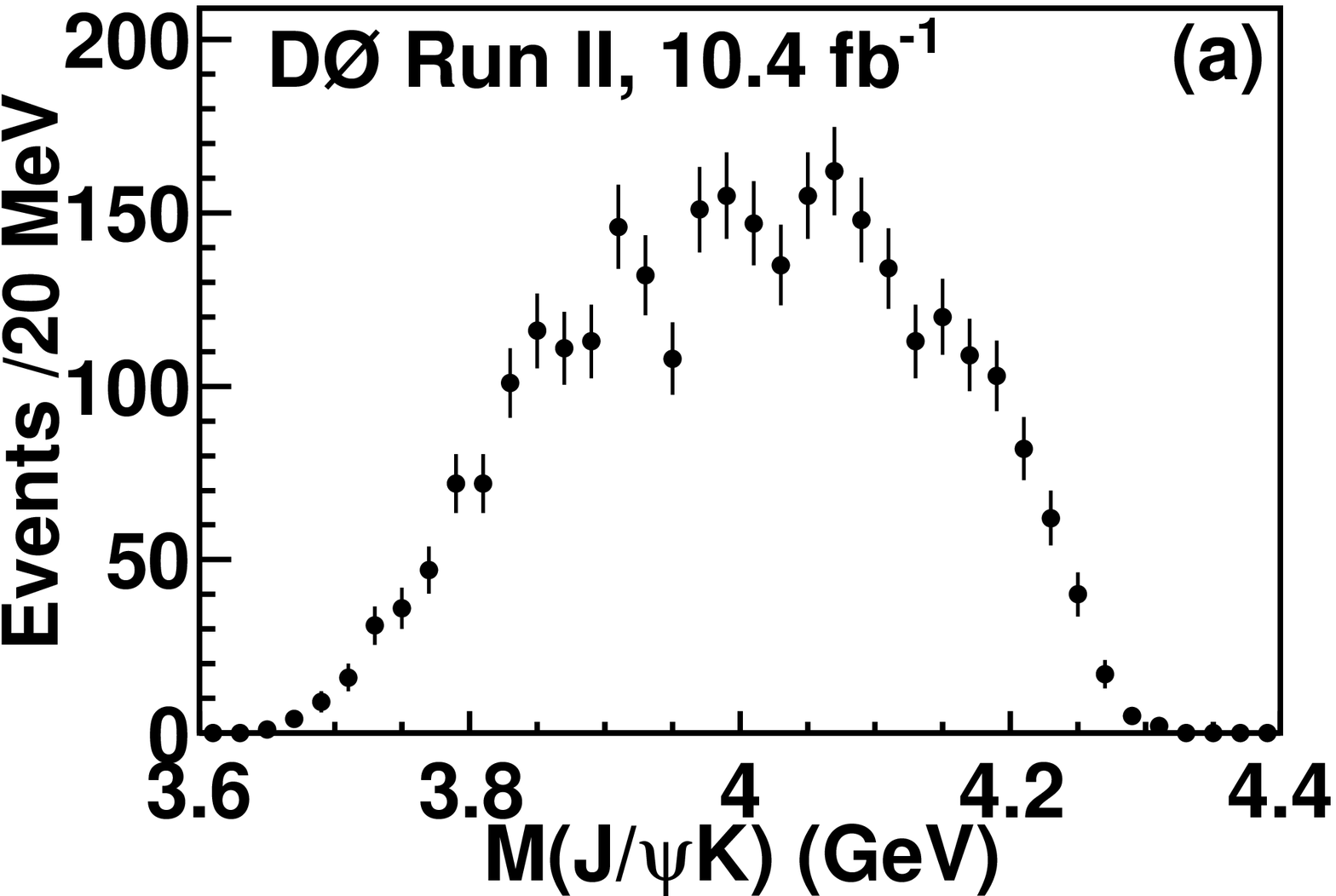}
  \includegraphics[width=0.80\columnwidth]{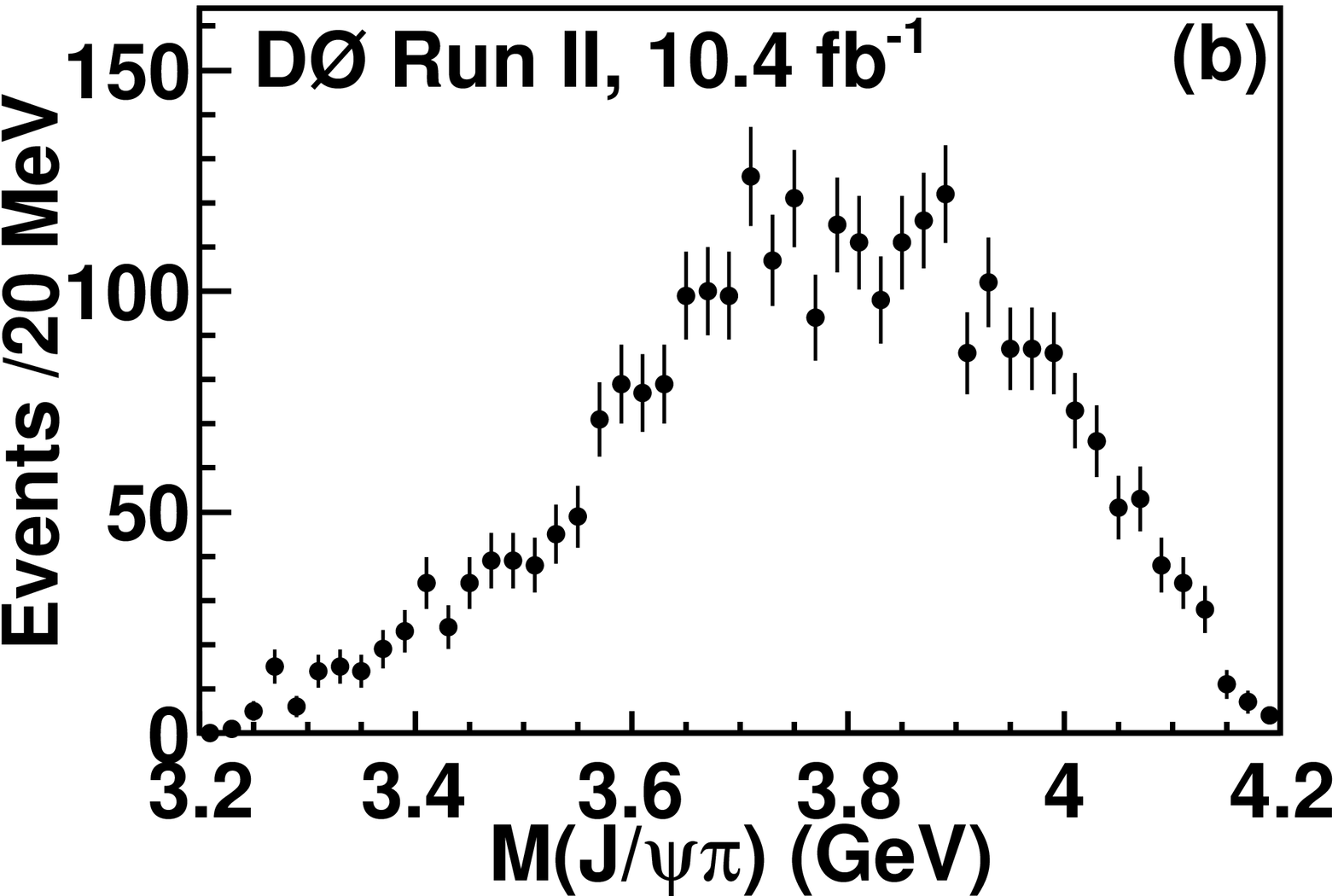}
  \caption{(a) Invariant mass distribution of  $J/\psi K$ pairs after the mass requirements $5.23<M(B^+)<5.33$~GeV and $1.005<M(K^+K^-)<1.035$~GeV. (b) Invariant mass of the same pairs under the  $J/\psi \pi$ hypothesis. 
}
   \label{fig:projy}
\end{figure}

We examine combinations of $J/\psi$ with one, two, or three charged particles, searching for structures that would affect the analysis of the $J/\psi \phi$ distribution. There are multiple reasons for this study: (i) a resonance in a subsystem may create an enhancement in the $M(J/\psi \phi)$ distribution leading to a false signal, (ii) identifying resonances and applying appropriate mass restrictions to eliminate their effects would reduce background, (iii) finding a resonance and fitting its mass and width provides an in situ calibration of the mass and resolution for a given configuration.

Of particular concern is the new charged charmonium-like object, $Z(3900)^{\pm}$,  observed independently by the BESIII~\cite{Ablikim:2013mio} and Belle~\cite{Liu:2013dau} Collaborations in the $J/\psi \pi^+$  decay channel.  The distributions of $M(J/\psi K)$, where the $J/\psi$ is paired with the particle that is not associated with the $\phi$ decay in the reconstructed $B^+$ decay, is shown in Fig.~\ref{fig:projy}(a). No structures that would indicate resonances or reflections of other decays are observed. 
 The mass distribution for the same pair under the pion mass assignment, shown in Fig.~\ref{fig:projy}(b), is also structureless.

\begin{figure}[h!tb]
  \centering
   \includegraphics[width=0.80\columnwidth]{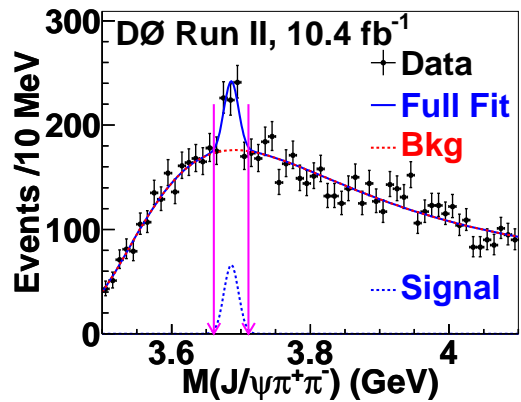}
  \caption{(color online)  Invariant mass distribution $M(J/\psi \pi^+ \pi^-)$ before the $\psi(2S)$ veto.
The fit assumes a Gaussian $\psi(2S)$ signal  with the mean mass set to the world-average value
~\cite{pdg2012}, and a free resolution parameter. The arrows indicate the $\pm2.5$ standard deviation range excluded from the analysis. The background  is described by a product of a Landau function and an exponential. 
}
   \label{fig:reflections}
\end{figure}

The  $M(J/\psi \pi^+ \pi^-)$ distribution,   before application of the $\psi(2S)$ veto, is shown in Fig.~\ref{fig:reflections}. In the fit the resonance mass is set at the world-average value  
of $\psi(2S)$ mass~\cite{pdg2012}. A fit with the mass allowed to vary gives the value consistent with the world-average value. The resolution of 10 MeV is consistent with simulations.
 There are no enhancements other than the  $\psi(2S)$ meson peak that we remove from the sample.

\begin{figure}[h!tb]
  \centering
   \includegraphics[width=0.80\columnwidth]{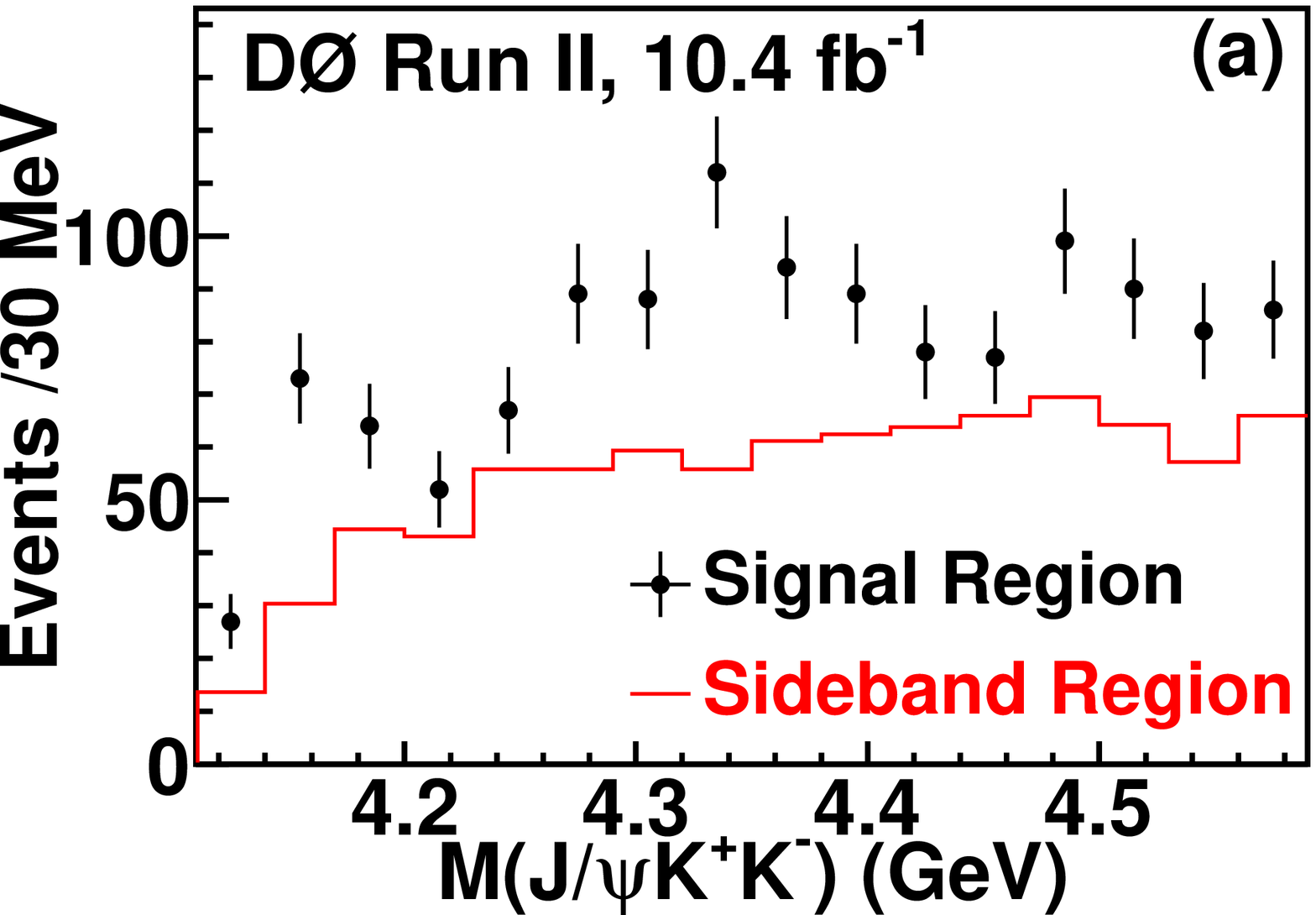}
  \includegraphics[width=0.80\columnwidth]{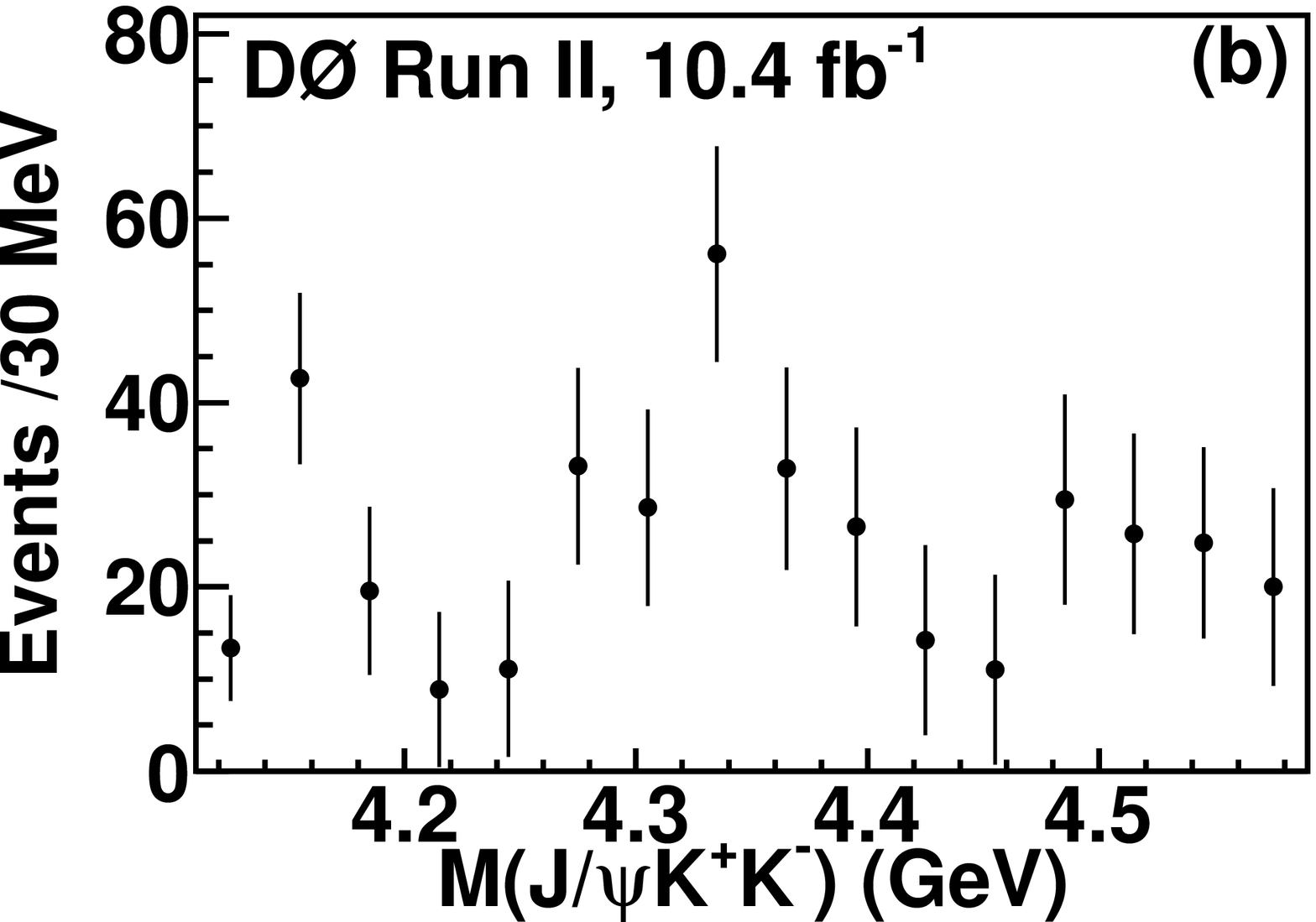}
  \caption{(color online)  Invariant mass distribution $M(J/\psi \phi)$   after the mass requirements $5.23<M(B^+)<5.33$~GeV and $1.005<M(K^+K^-) <1.035$~GeV. The background is estimated from the $B^+$ sidebands. (b) Difference between the distributions of the signal and normalized background. 
}
   \label{fig:ycomp}
\end{figure}

Figure~\ref{fig:ycomp} shows the invariant mass distribution of the $J/\psi \phi$ candidates within the $B^+$ and $\phi$ mass windows. Overlaid is the background distribution estimated from the $B^+$ sidebands. An enhancement at low masses and a broader enhancement near 4.3 GeV are seen, consistent with the CDF~\cite{Aaltonen:2009at} and CMS~\cite{cms} results. 

Small statistics  and high background  do not allow a detailed two-dimensional analysis of the three-body $B^+$ decay.
We therefore focus on the one-dimensional projection of data on  $M(J/\psi \phi)$.
In the search for the particular state $X(4140)$, we define the allowed region for a possible resonance mass
as  $M(J/\psi \phi)<4.20$~GeV, well above the $X(4140)$ mass value, taking into account
our mass resolution. 
 There are 80 events in this region. According to  ensemble tests,
with a large number of  pseudo-experiments with the same signal and backgroud statistics,
and assuming a direct three-body $B^+$ decay,
the probability of the phase space fluctuation to this value is $8\times10^{-4}$.

We divide the sample into  30 MeV wide intervals in $M(J/\psi \phi)$  from 4.11 to 4.59 GeV and fit the subsamples for the number of events of the $B^+$ decay (the bin centered at  $M(J/\psi \phi)$ =4.155 GeV is further divided into two parts). In the fits,  we constrain the $B^+$ mean mass, as well as the parameters describing the background shape, to the values obtained in the overall fit shown in Fig.~\ref{fig:ball}. According to simulations, the $B^+$ mass resolution varies from 20 MeV for  $M(J/\psi \phi)<4.3$ GeV to 17 MeV for  $M(J/\psi \phi)>4.5$ GeV. This variation is taken into account in the fits.

\begin{figure}[h]
  \centering
   \includegraphics[width=0.80\columnwidth]{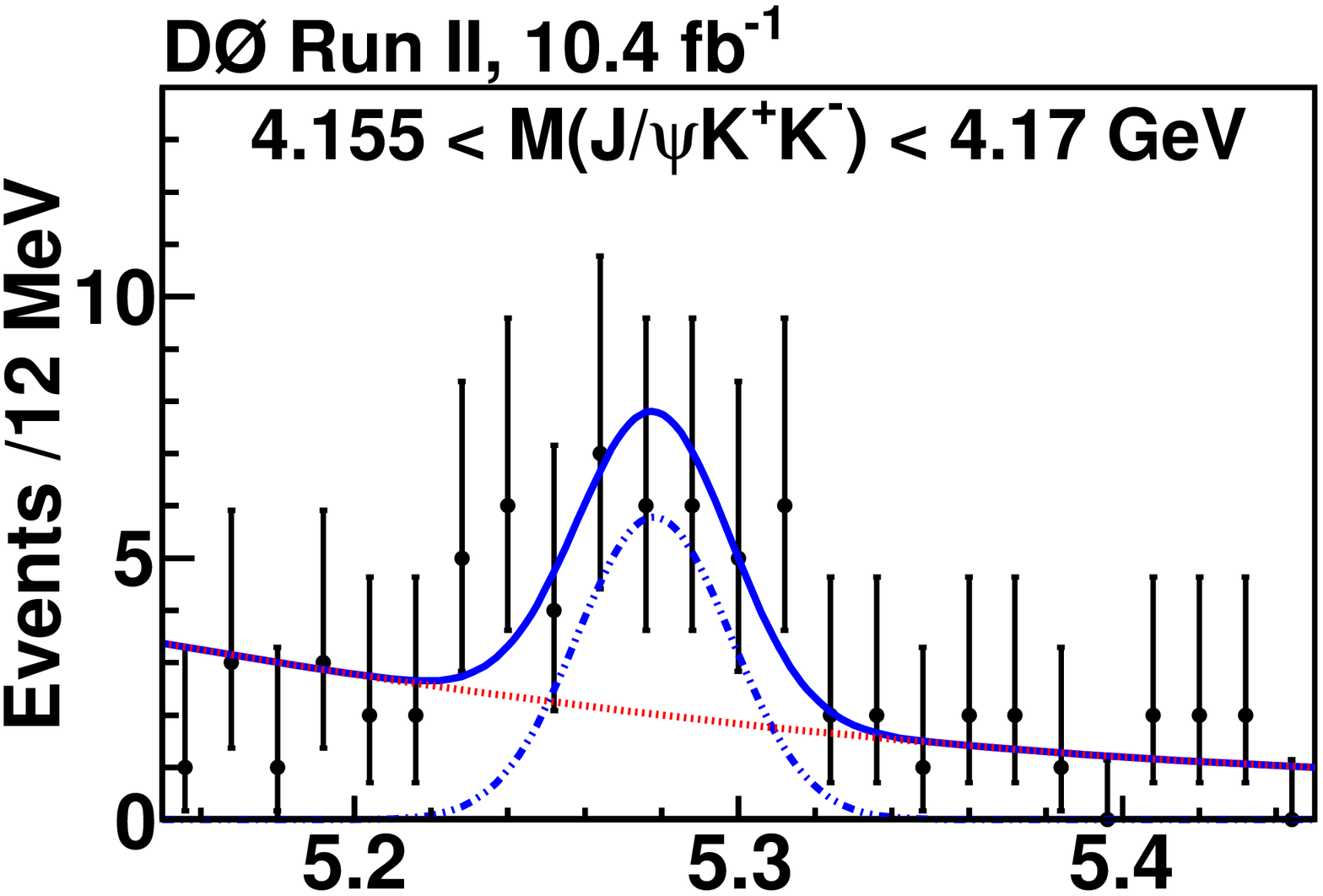}
   \includegraphics[width=0.80\columnwidth]{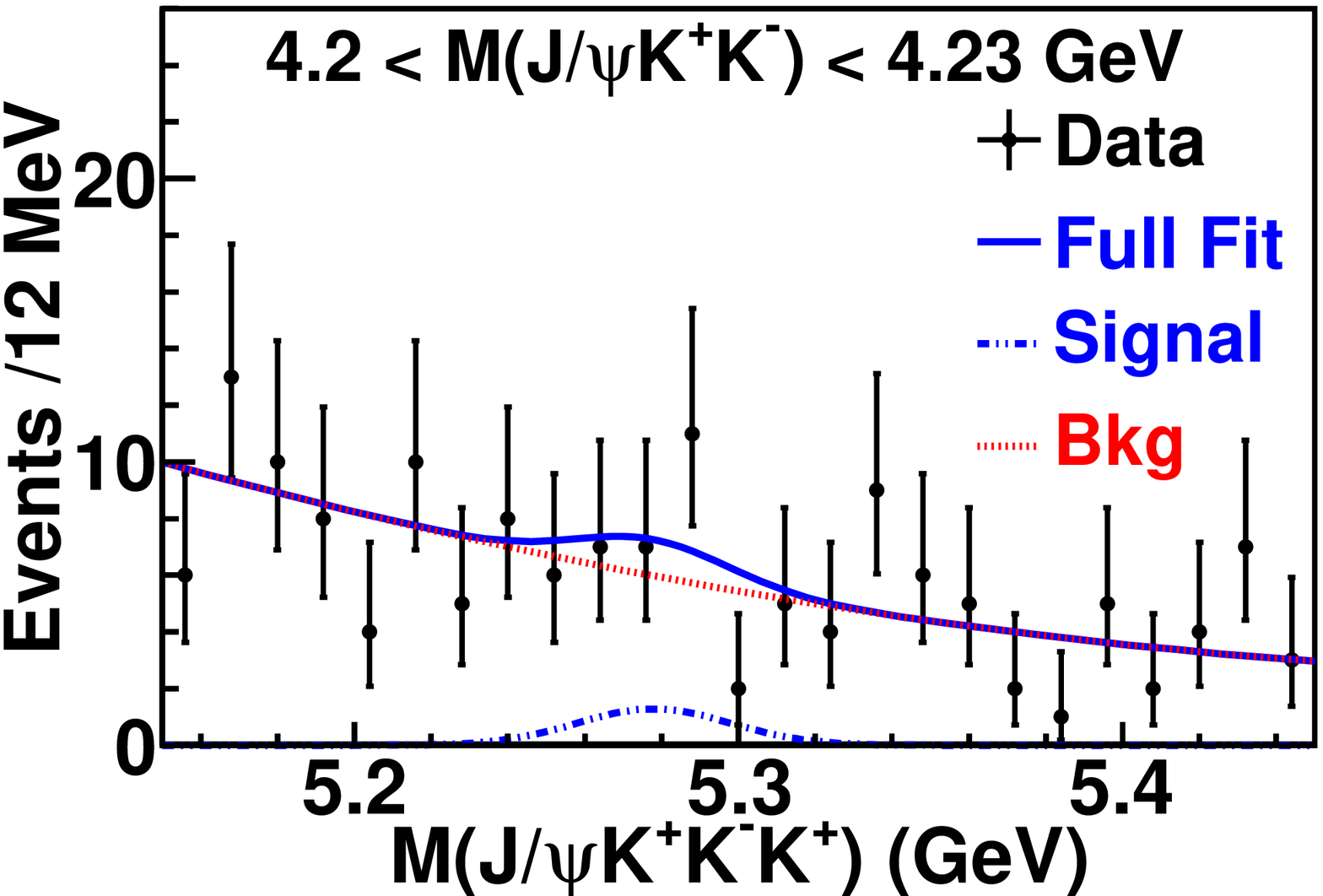}

  \caption{(color online) Invariant mass distributions of $B^+ \rightarrow J/\psi \phi K^+$ candidates in two selected intervals
of  $M(J/\psi \phi)$. Superimposed are the fits  of a Gaussian signal (solid blue lines) with a second-order Chebyshev polynomial   background (dashed red lines), with the signal and background shape parameters constrained to the results of the fit in
Fig.~\ref{fig:ball}, and allowing for the signal yield to vary.
}
   \label{fig:y16}
\end{figure}

\begin{figure}[htb]
  \centering
   \includegraphics[width=0.80\columnwidth]{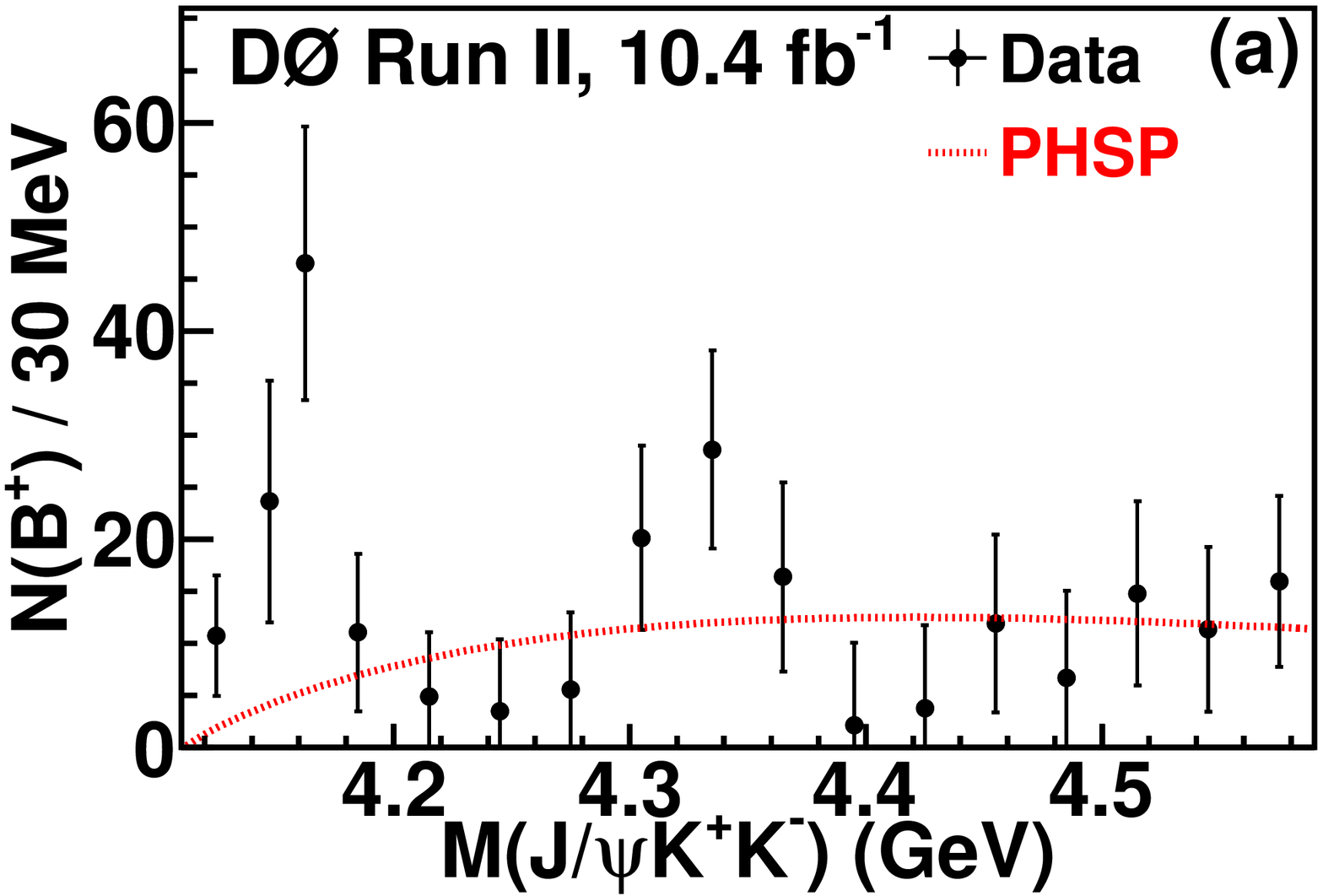}
   \includegraphics[width=0.80\columnwidth]{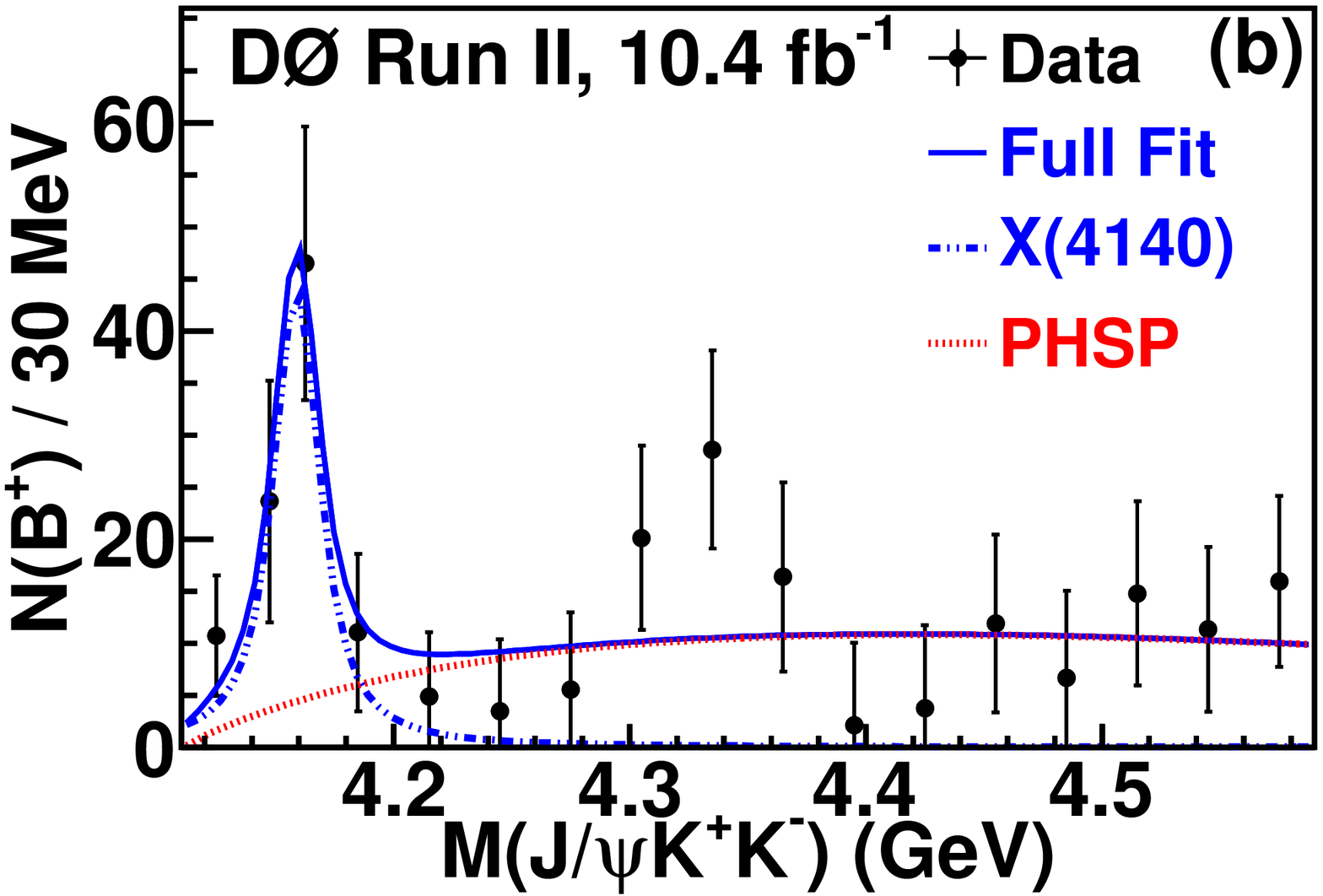}
   \includegraphics[width=0.80\columnwidth]{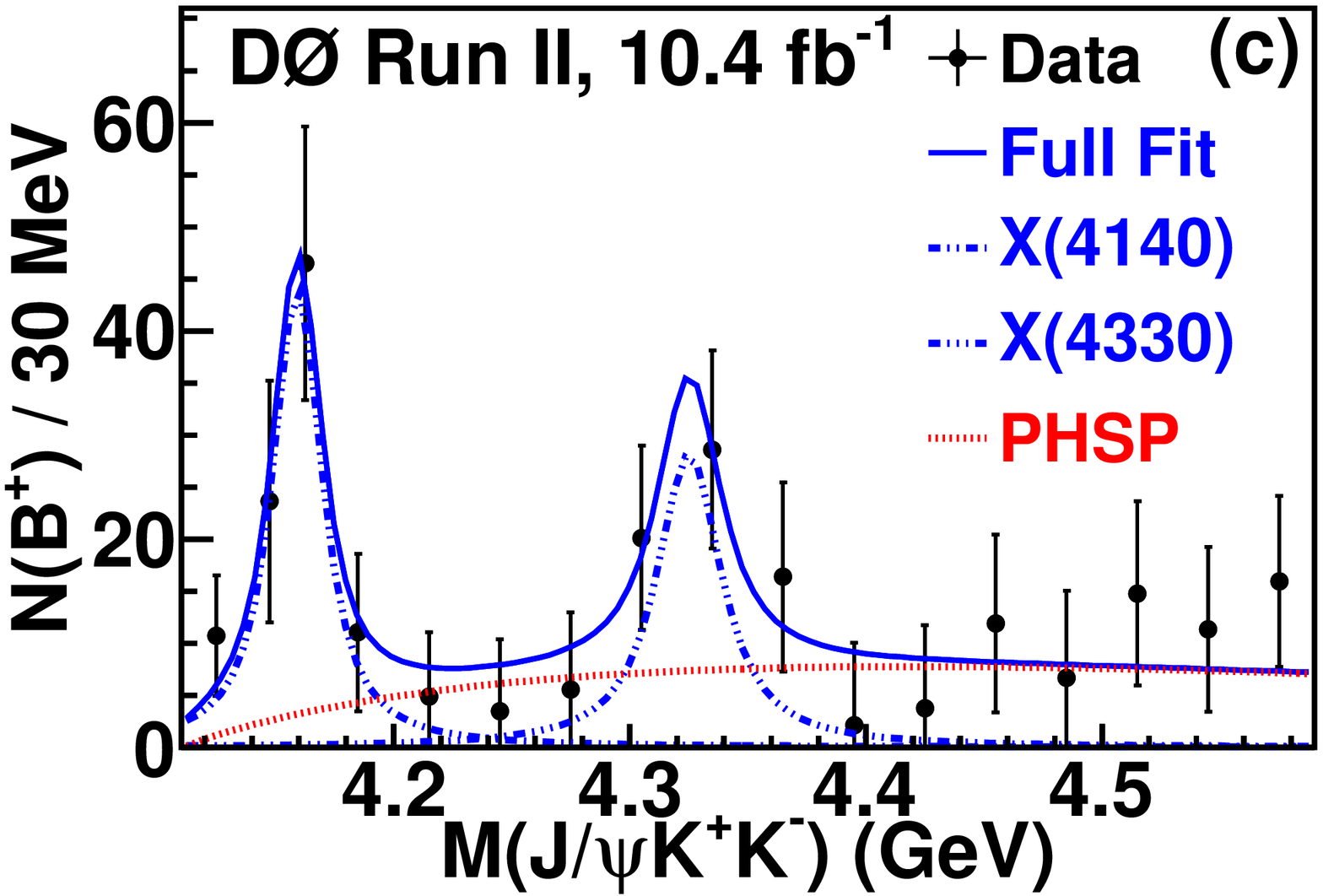}
  \caption{(color online) The $B^+ \rightarrow J/\psi \phi K^+$ signal yield per 30 MeV resulting
 from fits in 17 $M(J/\psi \phi)$ bins defined in the text,  corrected for acceptance. 
Note that the second and third bins have widths of 15 MeV, and the points are normalized to the counts per 30 MeV
 as the rest of the bins. (a) Fit allowing for no $J/\psi \phi$ resonance and assuming a three-body phase-space (PHSP)~\cite{pdg2012};  (b) allowing for a Breit-Wigner $X(4140)$ signal with an unconstrained mass and width and with a resolution of 4~MeV; (c) allowing for two Breit-Wigner resonances where the natural width of the second is set to 30 MeV.
The resonance contributions, the three-body phase-space contribution, and the total fit  are also shown.  }
   \label{fig:yfit}
\end{figure}

\begin{figure}[htb]
  \centering
   \includegraphics[width=0.80\columnwidth]{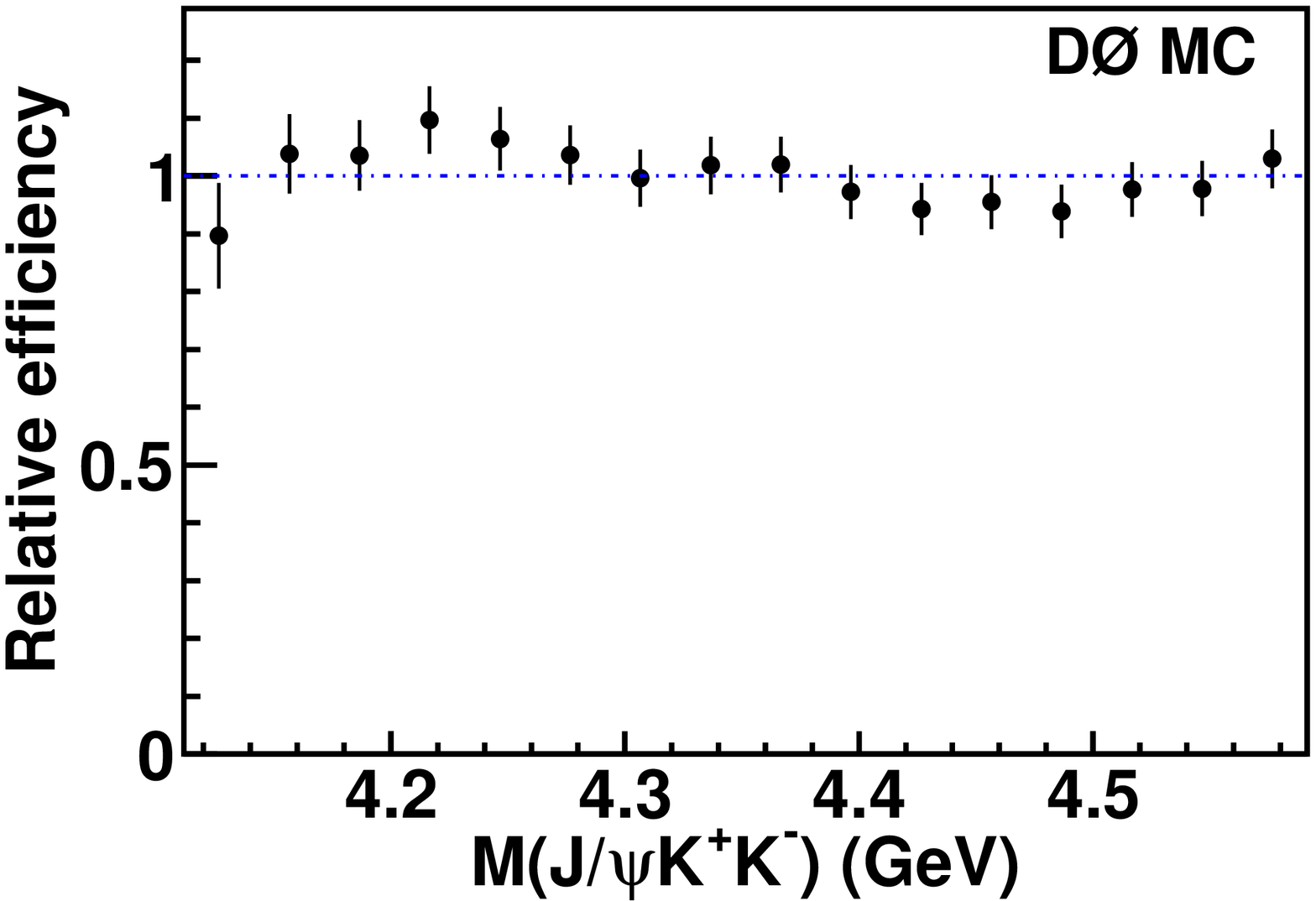}
  \caption{
Relative efficiency as a fucntion of of  $M(J/\psi \phi)$ due to  kinematic acceptance, reconstruction efficiency, and resolution.}
   \label{fig:eff}
\end{figure}

Two examples of the  distributions are shown in Fig.~\ref{fig:y16}. The resulting $B^+$ yield per 30 MeV  as a function of  $M(J/\psi \phi)$, corrected for efficiency, is shown in Fig.~\ref{fig:yfit}.  The relative efficiency as a function of $J/\psi \phi$ mass is obtained by comparing the reconstructed spectrum from a full detector simulation with the three-body phase space distribution. 
 The  efficiency correction includes effects of the kinematic acceptance, as well as the reconstruction efficiency, the resolution, and the candidate selection efficiency. As shown in Fig.~\ref{fig:eff}, the efficiency is fairly uniform, with bin-to-bin variations within 10\%.

 To estimate the significance of the threshold structure, we perform a binned least-squares fit of the $B^+$ yield to a sum of a resonance and a phase-space continuum template. We assume a relativistic Breit-Wigner signal shape, with mass and width allowed to vary, convoluted with the detector resolution of 4~MeV from simulations.  From the fit,   shown in Fig.~\ref{fig:yfit}(b), we obtain $52\pm 19 {\rm \thinspace (stat)}$ signal events out of the total of $250\pm36$ events.
The statistical significance of the structure, estimated from the $\chi^2$ difference  with and without a resonant component, $\Delta \chi^2 = 14.7$ for 3 degrees of freedom, is 3.1 standard deviations. The fitted mass of this state is $4159.0\pm4.3 {\rm \thinspace (stat)}$~MeV and the width is $19.9\pm12.6 {\rm \thinspace (stat)}$~MeV. We identify this structure with $X(4140)$ and we find that the quasi-two body decay $B^+ \rightarrow X(4140) K^+$ constitutes ($21 \pm 8 {\rm \thinspace (stat)}$)\% of the  $B^+ \rightarrow J/\psi \phi K^+$ decay rate.  The data also support the presence of a structure around 4300~MeV, however they do not allow a stable fit with an unconstrained width. When a second resonance is allowed
by setting the natural width to 30 MeV, consistent with the CDF data, the fit as shown in  Fig.~\ref{fig:yfit}(c) returns $47\pm 20$ events at an invariant  mass of $4328.5 \pm 12.0$ MeV.


The  $X(4140)$ mass and width measurements and the relative branching fraction are subject to systematic uncertainties associated with the precision of  the $B^+$  mass  measurement, with the $J/\psi \phi$ mass resolution in the vicinity of $X(4140)$, and  with the variation of the reconstruction efficiency with $M(J/\psi \phi)$.
 To estimate these uncertainties, we perform alternative fits applying more restrictive event selection criteria, using a different bin size, and fitting the net mass distribution of $J/\psi \phi$ pairs
coming from $B^+$ decay obtained by subtracting the properly normalized background from the sideband region. 
In addition, we consider the following variations of the $B^+$ mass fits in $M(J/\psi \phi)$ intervals:
We vary the $B^+$ mean mass by its uncertainty of $\pm3$~MeV,
vary the $B^+$ mass resolution by its uncertainty of $\pm 1$~MeV,
vary background parameters within their uncertainties and use a third-order Chebyshev polynomial
in the fit to the background.

In the nominal fits of the signal yield as a function of $M(J/\psi \phi)$,
we use the  $J/\psi \phi$ mass resolution of 4 MeV as obtained in simulations. 
For decay processes with a similar topology, $\psi(2S) \rightarrow J/\psi \pi^+ \pi^-$ and $X(3872) \rightarrow J/\psi \pi^+ \pi^-$,
the measured mass resolutions are $9.9 \pm 0.3$~MeV, and $15.9\pm 3.2$~MeV, respectively. Both are in a good agreement with simulations.
Since the mass resolution is better for lower values of the kinetic energy released in the decay,
 the resolution for the structures under study is not larger than that for the
 $\psi(2S)$ decay.
We repeat the analysis using the value of 10~MeV. 
The change in the resolution does not affect the results for the resonance mass and yield but
 it reduces its width. We assign an asymmetric uncertainty of $-8$~MeV due to this effect.

We vary the efficiency dependence on $M(J/\psi \phi)$ within the statistical uncertainties.
In alternative fits, we use a relative efficiency that is independent of $M(J/\psi \phi)$ and also
a relative efficiency that drops to the value of 0.8 (instead of the default value of  $0.9\pm 0.1$) at the $M(J/\psi \phi)$ threshold.

 From the results of the alternative fits we estimate the systematic uncertainties on the $X(4140)$ mass and width to be
$\pm 4$~MeV and $^{+1}_{-8}$~MeV, and the systematic uncertainty of the relative branching fraction to be $\pm 4$\%.
We estimate the uncertainty in the $J/\psi \phi$ mass scale from the vertex reconstruction by comparing $M_{\rm def}=M(J/\psi \phi$) value used in this analysis with the alternative estimate, obtained from the mass difference $M_{\rm alt} = M(\mu^+ \mu^- K^+K^-) -  M(\mu^+ \mu^-)  +M(J/\psi)$.
 The difference $\Delta M=  M_{\rm def} -  M_{\rm alt}$ is on average  $1.3$~MeV, and the $RMS$ is 5.2~MeV. We conclude that there is no significant mass bias due to the vertexing procedure and we conservatively assign a systematic uncertainty of $\pm5.2$~MeV due to the uncertainty in the $J/\psi \phi$ mass scale.
The statistical significance of the $X(4140)$ signal  is larger than 3 standard deviations in all alternative fits. 
A search  conducted in the entire mass range
$(4.11, 4.59)$~GeV, ignoring the prior observation of $X(4140)$, would result in the signal significance reduced
due to the ``look elsewhere effect''~\cite{lee} by the  trial factor of 5 to 2.6 standard deviations.

In summary, in the decay $B^+ \rightarrow J/\psi \phi K^+$, we find a threshold enhancement in the $J/\psi \phi$ mass distribution consistent with the $X(4140)$ state with a statistical significance of 3.1 standard deviations.
The data can also accommodate a second structure, at $M_2=4328.5\pm 12.0$~MeV, consistent with Ref.~\cite{Shen:2009vs}.
 The measured invariant mass of the lower-mass peak  is $4159.0\pm4.3 {\rm \thinspace (stat)} \pm 6.6 {\rm \thinspace (syst)}$ MeV and  the measured width is $19.9\pm12.6 {\rm \thinspace (stat)}^{+1}_{-8}  {\rm \thinspace (syst)}$~MeV. The relative branching fraction ${\cal B}_{\rm rel}= {\cal B}(B^+ \rightarrow X(4140) K^+)/ {\cal B}(B^+ \rightarrow J/\psi \phi K^+)$ (for $M(J/\psi \phi) <4.59$~GeV) is measured to be ($21 \pm 8 {\rm \thinspace (stat)} \pm 4 {\rm \thinspace (syst)}$)\%. 
Our results support the existence of the $X(4140)$ resonance.

\input acknowledgement.tex   

\bibliographystyle{atlasnote}
\bibliography{yrc}

\end{document}

%% file: author_list.tex
\affiliation{LAFEX, Centro Brasileiro de Pesquisas F\'{i}sicas, Rio de Janeiro, Brazil}
\affiliation{Universidade do Estado do Rio de Janeiro, Rio de Janeiro, Brazil}
\affiliation{Universidade Federal do ABC, Santo Andr\'e, Brazil}
\affiliation{University of Science and Technology of China, Hefei, People's Republic of China}
\affiliation{Universidad de los Andes, Bogot\'a, Colombia}
\affiliation{Charles University, Faculty of Mathematics and Physics, Center for Particle Physics, Prague, Czech Republic}
\affiliation{Czech Technical University in Prague, Prague, Czech Republic}
\affiliation{Institute of Physics, Academy of Sciences of the Czech Republic, Prague, Czech Republic}
\affiliation{Universidad San Francisco de Quito, Quito, Ecuador}
\affiliation{LPC, Universit\'e Blaise Pascal, CNRS/IN2P3, Clermont, France}
\affiliation{LPSC, Universit\'e Joseph Fourier Grenoble 1, CNRS/IN2P3, Institut National Polytechnique de Grenoble, Grenoble, France}
\affiliation{CPPM, Aix-Marseille Universit\'e, CNRS/IN2P3, Marseille, France}
\affiliation{LAL, Universit\'e Paris-Sud, CNRS/IN2P3, Orsay, France}
\affiliation{LPNHE, Universit\'es Paris VI and VII, CNRS/IN2P3, Paris, France}
\affiliation{CEA, Irfu, SPP, Saclay, France}
\affiliation{IPHC, Universit\'e de Strasbourg, CNRS/IN2P3, Strasbourg, France}
\affiliation{IPNL, Universit\'e Lyon 1, CNRS/IN2P3, Villeurbanne, France and Universit\'e de Lyon, Lyon, France}
\affiliation{III. Physikalisches Institut A, RWTH Aachen University, Aachen, Germany}
\affiliation{Physikalisches Institut, Universit\"at Freiburg, Freiburg, Germany}
\affiliation{II. Physikalisches Institut, Georg-August-Universit\"at G\"ottingen, G\"ottingen, Germany}
\affiliation{Institut f\"ur Physik, Universit\"at Mainz, Mainz, Germany}
\affiliation{Ludwig-Maximilians-Universit\"at M\"unchen, M\"unchen, Germany}
\affiliation{Panjab University, Chandigarh, India}
\affiliation{Delhi University, Delhi, India}
\affiliation{Tata Institute of Fundamental Research, Mumbai, India}
\affiliation{University College Dublin, Dublin, Ireland}
\affiliation{Korea Detector Laboratory, Korea University, Seoul, Korea}
\affiliation{CINVESTAV, Mexico City, Mexico}
\affiliation{Nikhef, Science Park, Amsterdam, the Netherlands}
\affiliation{Radboud University Nijmegen, Nijmegen, the Netherlands}
\affiliation{Joint Institute for Nuclear Research, Dubna, Russia}
\affiliation{Institute for Theoretical and Experimental Physics, Moscow, Russia}
\affiliation{Moscow State University, Moscow, Russia}
\affiliation{Institute for High Energy Physics, Protvino, Russia}
\affiliation{Petersburg Nuclear Physics Institute, St. Petersburg, Russia}
\affiliation{Instituci\'{o} Catalana de Recerca i Estudis Avan\c{c}ats (ICREA) and Institut de F\'{i}sica d'Altes Energies (IFAE), Barcelona, Spain}
\affiliation{Uppsala University, Uppsala, Sweden}
\affiliation{Lancaster University, Lancaster LA1 4YB, United Kingdom}
\affiliation{Imperial College London, London SW7 2AZ, United Kingdom}
\affiliation{The University of Manchester, Manchester M13 9PL, United Kingdom}
\affiliation{University of Arizona, Tucson, Arizona 85721, USA}
\affiliation{University of California Riverside, Riverside, California 92521, USA}
\affiliation{Florida State University, Tallahassee, Florida 32306, USA}
\affiliation{Fermi National Accelerator Laboratory, Batavia, Illinois 60510, USA}
\affiliation{University of Illinois at Chicago, Chicago, Illinois 60607, USA}
\affiliation{Northern Illinois University, DeKalb, Illinois 60115, USA}
\affiliation{Northwestern University, Evanston, Illinois 60208, USA}
\affiliation{Indiana University, Bloomington, Indiana 47405, USA}
\affiliation{Purdue University Calumet, Hammond, Indiana 46323, USA}
\affiliation{University of Notre Dame, Notre Dame, Indiana 46556, USA}
\affiliation{Iowa State University, Ames, Iowa 50011, USA}
\affiliation{University of Kansas, Lawrence, Kansas 66045, USA}
\affiliation{Louisiana Tech University, Ruston, Louisiana 71272, USA}
\affiliation{Northeastern University, Boston, Massachusetts 02115, USA}
\affiliation{University of Michigan, Ann Arbor, Michigan 48109, USA}
\affiliation{Michigan State University, East Lansing, Michigan 48824, USA}
\affiliation{University of Mississippi, University, Mississippi 38677, USA}
\affiliation{University of Nebraska, Lincoln, Nebraska 68588, USA}
\affiliation{Rutgers University, Piscataway, New Jersey 08855, USA}
\affiliation{Princeton University, Princeton, New Jersey 08544, USA}
\affiliation{State University of New York, Buffalo, New York 14260, USA}
\affiliation{University of Rochester, Rochester, New York 14627, USA}
\affiliation{State University of New York, Stony Brook, New York 11794, USA}
\affiliation{Brookhaven National Laboratory, Upton, New York 11973, USA}
\affiliation{Langston University, Langston, Oklahoma 73050, USA}
\affiliation{University of Oklahoma, Norman, Oklahoma 73019, USA}
\affiliation{Oklahoma State University, Stillwater, Oklahoma 74078, USA}
\affiliation{Brown University, Providence, Rhode Island 02912, USA}
\affiliation{University of Texas, Arlington, Texas 76019, USA}
\affiliation{Southern Methodist University, Dallas, Texas 75275, USA}
\affiliation{Rice University, Houston, Texas 77005, USA}
\affiliation{University of Virginia, Charlottesville, Virginia 22904, USA}
\affiliation{University of Washington, Seattle, Washington 98195, USA}
\author{V.M.~Abazov} \affiliation{Joint Institute for Nuclear Research, Dubna, Russia}
\author{B.~Abbott} \affiliation{University of Oklahoma, Norman, Oklahoma 73019, USA}
\author{B.S.~Acharya} \affiliation{Tata Institute of Fundamental Research, Mumbai, India}
\author{M.~Adams} \affiliation{University of Illinois at Chicago, Chicago, Illinois 60607, USA}
\author{T.~Adams} \affiliation{Florida State University, Tallahassee, Florida 32306, USA}
\author{J.P.~Agnew} \affiliation{The University of Manchester, Manchester M13 9PL, United Kingdom}
\author{G.D.~Alexeev} \affiliation{Joint Institute for Nuclear Research, Dubna, Russia}
\author{G.~Alkhazov} \affiliation{Petersburg Nuclear Physics Institute, St. Petersburg, Russia}
\author{A.~Alton$^{a}$} \affiliation{University of Michigan, Ann Arbor, Michigan 48109, USA}
\author{A.~Askew} \affiliation{Florida State University, Tallahassee, Florida 32306, USA}
% ...
%
%
\collaboration{The D0 Collaboration\footnote{with visitors from
$^{a}$Augustana College, Sioux Falls, SD, USA,
$^{b}$The University of Liverpool, Liverpool, UK,
$^{c}$DESY, Hamburg, Germany,
$^{d}$Universidad Michoacana de San Nicolas de Hidalgo, Morelia, Mexico
$^{e}$SLAC, Menlo Park, CA, USA,
$^{f}$University College London, London, UK,
$^{g}$Centro de Investigacion en Computacion - IPN, Mexico City, Mexico,
$^{h}$Universidade Estadual Paulista, S\~ao Paulo, Brazil,
$^{i}$Karlsruher Institut f\"ur Technologie (KIT) - Steinbuch Centre for Computing (SCC)
and
$^{j}$Office of Science, U.S. Department of Energy, Washington, D.C. 20585, USA.
}} \noaffiliation
\vskip 0.25cm

%% file: acknowledgement.tex
%
We thank the staffs at Fermilab and collaborating institutions,
and acknowledge support from the
DOE and NSF (USA);
CEA and CNRS/IN2P3 (France);
MON, NRC KI and RFBR (Russia);
CNPq, FAPERJ, FAPESP and FUNDUNESP (Brazil);
DAE and DST (India);
Colciencias (Colombia);
CONACyT (Mexico);
NRF (Korea);
FOM (The Netherlands);
STFC and the Royal Society (United Kingdom);
MSMT and GACR (Czech Republic);
BMBF and DFG (Germany);
SFI (Ireland);
The Swedish Research Council (Sweden);
and
CAS and CNSF (China).